\documentclass[pra,aps,onecolumn,showpacs,epsf]{revtex4}
\usepackage[english]{babel}
\usepackage{epsfig}
\usepackage{graphicx}
\usepackage{amsmath}
\usepackage{amssymb}
\usepackage{subfigure}
\usepackage{bm}

\begin{document}

\title{Photoconductivity of CdS-CdSe granular films: influence of
microstructure}
\author{A. S. Meshkov$^{1}$, E. F. Ostretsov$^{2}$, W. V. Pogosov$^{1}$, I.
A. Ryzhikov$^{1}$, Yu. V. Trofimov$^{2}$}
\affiliation{$^{1}$Institute for Theoretical and Applied Electrodynamics, Russian Academy
of Sciences, Izhorskaya 13, 125412 Moscow, Russia}
\affiliation{$^{2}$SE "Center of LED and Optoelectronic Technologies of National Academy
of Sciences of Belarus", Logoiski trakt str. 22, 220090 Minsk, Belarus}

\begin{abstract}
We study experimentally the photoresistance of CdS-CdSe sintered granular
films obtained by the screen printing method. We mostly focus on the
dependences of photoresistance on film's microstructure, which varies with
changing heat-treatment conditions. The minimum photoresistance is found for
samples with compact packing of individual grains, which nevertheless are
separated by gaps. Such a microstructure is typical for films heat-treated
during an optimal time of 30 min at the temperature of 823 K. In order to
understand whether the dominant mechanism of charge transfer is identical
with the one in monocrystals, we perform temperature measurements of
photoresistance. Corresponding curves have the same nonmonotonic shape as in
CdSe monocrystals. Namely, photoresistance first increases with the growth
of temperature, and then starts to decrease. Thus we conclude that the basic
mechanism is also the same, as in monocrystals, and it is based on two types
of centers in the forbidden gap. We suggest that the optimal heat-treatment
time depends on two competing mechanisms: improvement of film's connectivity
and its oxidation. Photoresistance is also measured in vacuum and in helium
atmosphere, which suppress oxygen and water absorption/chemisorption at
intergrain boundaries. We demonstrate that this suppression decreases
photoresistance, especially at high temperatures.
\end{abstract}

\author{}
\maketitle

\section{Introduction}

Materials with granular structure are widely used in modern electronics and
radio engineering. These materials can be fabricated either by the
traditional screen printing method known for a long time or by using new
technologies, such as the direct forming of electronic devices by printers
and opal crystals creation by concretion of spherical microparticles from
water dredges. Modern technologies enable one to fabricate much more pure
and reproducible samples compared to the old ones. As a rule, by using these
methods materials with granular or polycrystalline internal structure are
obtained. In general, it is not always clear a priori how granularity
affects various electrophysical characteristics of a material \cite{BuFa}.
Therefore, electronic and optical properties of granular semiconductors
attract a lot of attention now.

Films made of CdS$\ $and CdSe are\ known for their high photoconductivity
and photosensitivity. Photoresistors based on these materials are widely
used in applications as photodetectors and optical couplers \cite{1,2,3,4}.
The investigations of photoconductive CdSe and CdS systems have a long
history, see e.g. Refs. \cite{Bube,Rose}. It was realized long time ago that
a model with the single type of recombination centers cannot account for
various electronic properties of CdSe and CdS crystals, such as
superlinearity of photoresistance, infrared and thermal quenching \cite{Rose}%
. Instead a model based on two types of states in the forbidden gap was
proposed \cite{Rose} (see also Ref. \cite{Stupp}, where even two more kinds
of centers were introduced). It was demonstrated \cite{Rose,Bube,Kind} that
photosensitivity of CdSe monocrystals can be as high as $10^{6}-10^{8}$.
Granular structure of thin films made of CdSe or CdS makes the physics of
these systems even more complicated \cite%
{BuFa,Bube,Petritz,Micheletti,Orton,Carbone} due to possible formation of
space charged regions inside individual crystallites and at boundaries.

In the recent paper \cite{Radio} some of us studied the influence of
heat-treatment conditions for CdS$_{1-x}$Se$_{x}$ granular films, obtained
by the screen printing method, on their photoresistance. It was found that
there is some optimal time for sample's heat-treatment (at fixed temperature
of heat-treatment), for which the photoresistance is minimized. The
light-to-dark current ratio for these films can be as large as 10$^{9}$.
This fact, together with the simple and low-cost method of film's
fabrication, make them quite attractive for technological applications. The
characteristic feature of optimally-prepared films is that neighboring
grains, from which they are build, fit each other on large contact areas. At
the same time, long time of heat-treatment results in the recrystallization
of grains and disappearance of spaces between them, so that the film's
connectivity is significantly improved compared to optimally-prepared films.
From these studies, it has remained unclear whether the leading mechanism of
charge transfer under the illumination is associated with intergrain
boundaries or it is identical with that for monocrystals of the same
chemical composition. It is also not evident why almost total disappearance
of gaps between grains is accompanied by the increase of photoresistance,
while it is expectable that gaps prevent grain-to-grain charge transfer.

The main goal of the present paper is to understand if the leading mechanism
of charge transfer in these films is different from the one in crystals. For
simplicity, we here restrict ourselves to CdS$_{0.2}$Se$_{0.8}$ films only.
In addition to the visual analysis of SEM images of film's microstructure,
we perform AFM studies of intergrain boundaries. We also measure temperature
dependences of photoresistance and compare the character of these quite
nontrivial dependences with the ones for monocrystals, which are known from
literature. Additionally, we perform measurements in vacuum, as well as in
the atmosphere of helium, which suppresses oxygen and water
absorption/chemisorption at grain boundaries, and compare the obtained
results with the ones for the atmosphere of air. These measurements indicate
that surfaces indeed play some role in the charge transfer under the
illumination, since absorbed oxygen and water increase photoresistance. We
finally make a conclusion that the leading mechanism of charge transfer in
the systems studied is most probably the same as in monocrystals, while the
existence of the optimal heat-treatment time is due to the competition
between the oxidation of grain boundaries during their heat-treatment, which
suppresses charge transfer, and improvement of film's connectivity, which
facilitates this process.

The paper is organized as follows. Section II deals with the sample
characterization. In Section III, we describe our experimental setup. In
Section IV, we present the results for the measurements of photoresistance
and discuss them. We conclude in Section V.

\section{Sample characterization}

We used CdS and CdSe powders as initial components for the paste, from which
samples are then fabricated. The initial powders of CdS\ and CdSe were
milled together. The prepared CdS-CdSe powder and the coupler were mixed in
a special barrel in order to obtain the paste. Propylene glycol was used as
a coupler. Through a stencil, this paste was deposited on the pyroceramics
substrate and then seasoned at room temperature for 0.5 hour. After that,
samples were dried for 1 hour at the temperature of 373 K in order to remove
the coupler. This raw material, represented by a film of 15-20 mkm
thickness, was heat-treated in the stove with a quasi-free air access
through the untight cup of the crucible. The time of heat-treatment for
different samples varied from 5 to 90 minutes, and the temperature - from
773 K to 873 K. Finally, samples were washed out in a bidistilled water and
dried at room temperature. By using this method, we have prepared more than
50 samples, which were differing from each other by chemical composition, as
well as by heat-treatment parameters, i.e., heat-treatment time and
temperature. The similar technology of sample fabrication was used by some
of us in Ref. \cite{14}.

\begin{figure}[tbp]
\begin{center}
\includegraphics[width=0.4\textwidth]{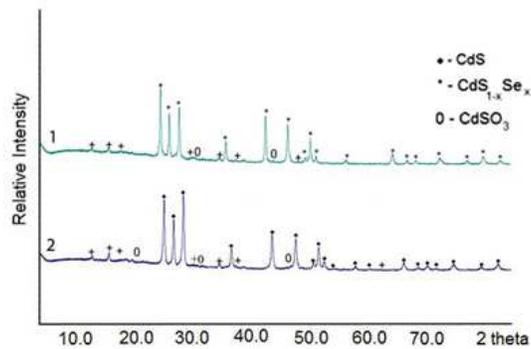}
\end{center}
\caption{XRD patterns of CdSe-CdS sample (curve 1) and standard CdS sample
(curve 2).}
\label{Fig1}
\end{figure}

During the heat-treatment process, small particles of the powder
corresponding to different chemical compositions, CdS or CdSe, merge into
grains, each grain containing large number of initial particles. As we
found, these grains consist of a solid CdS-CdSe solution. In order to prove
the fact that the solid solution is indeed formed, we performed x-ray
analysis of fabricated films. The typical measured XRD patterns for samples
are presented in Fig. 1. Positions of narrow peaks on XRD patterns do
correspond to the solid solution. The shift in the positions of these peaks
depends on the ratio of initial components (CdS and CdSe). Note that a trace
quantity of oxide CdSO$_{3}$ was also detected in our samples.

\begin{figure}[tbp]
\begin{center}
\centering
\subfigure{\includegraphics[width=0.3\textwidth]{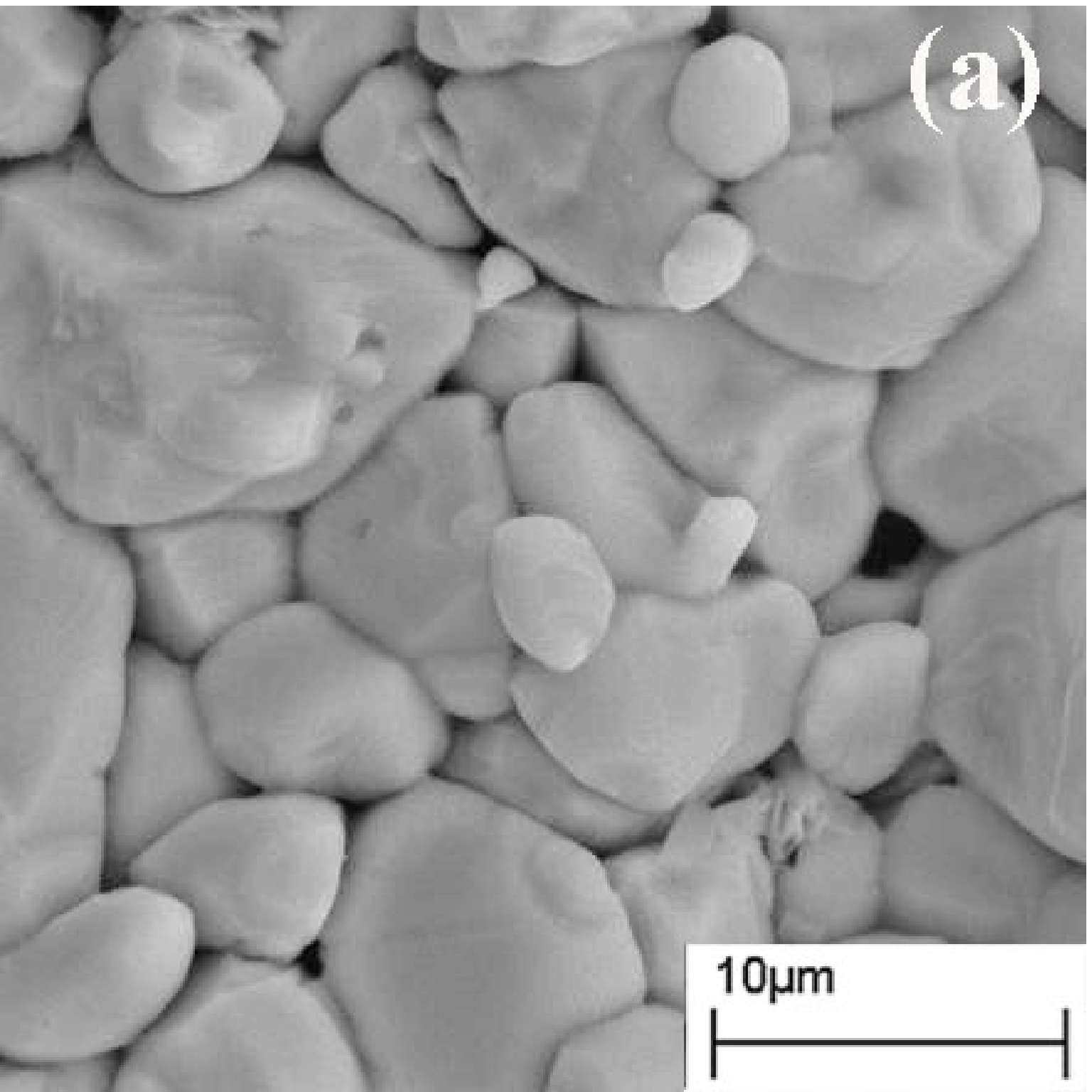}}\\[0pt]
\subfigure{\includegraphics[width=0.3\textwidth]{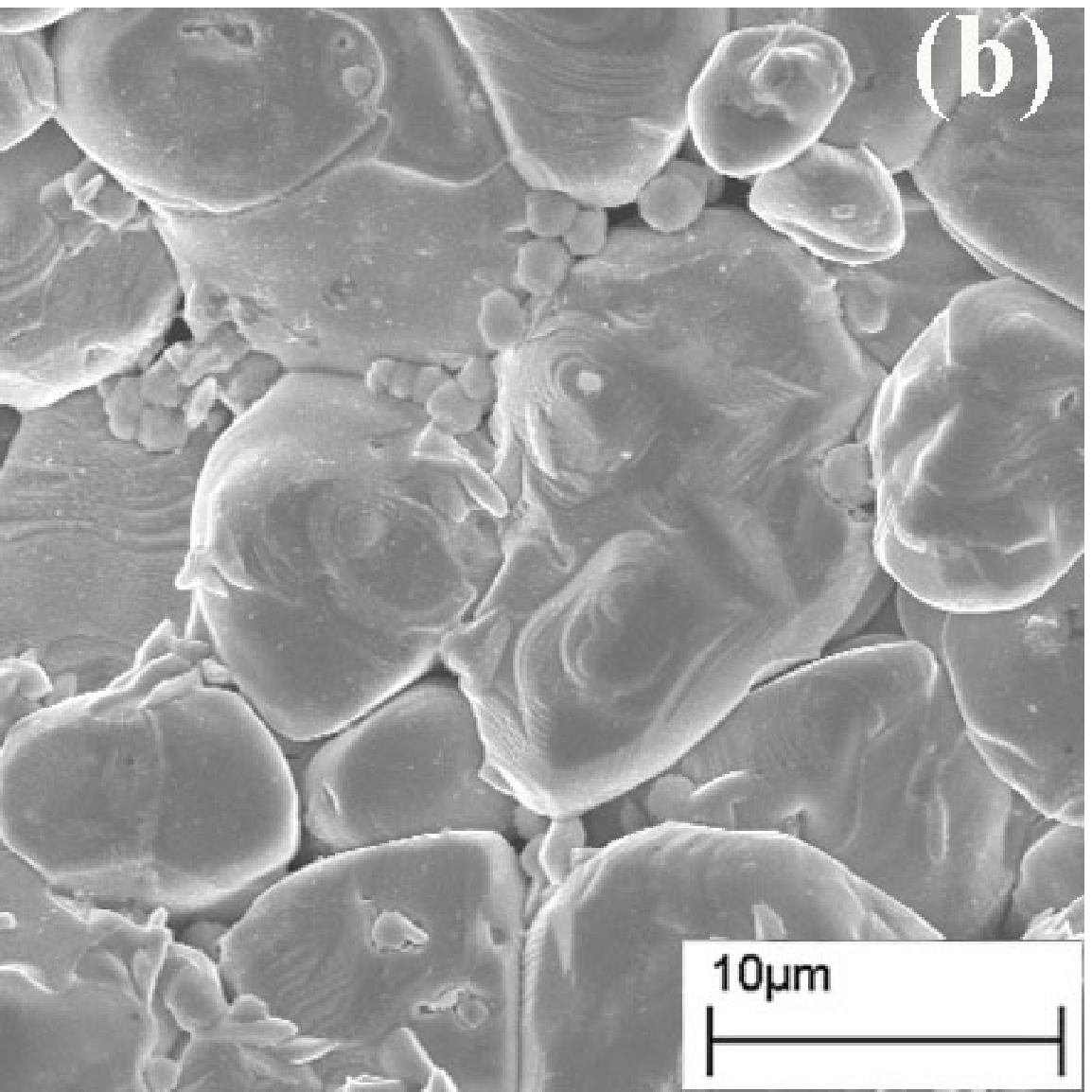}} %
\subfigure{\includegraphics[width=0.3\textwidth]{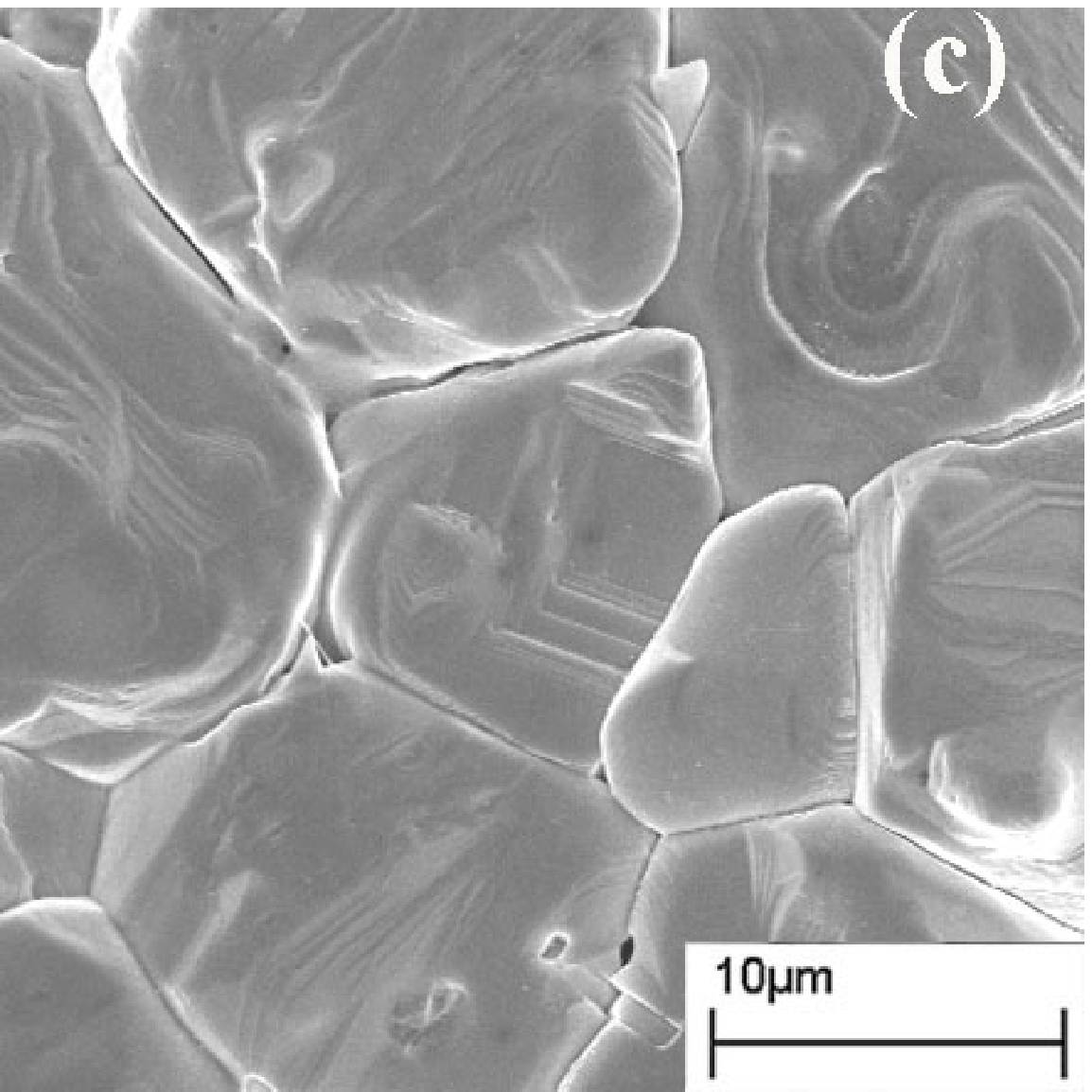}}
\end{center}
\caption{Micrographs of granular CdS$_{0.8}$Se$_{0.2}$ films fabricated with
different times of heat-treatment at fixed temperature $823$ $K$: 5 minutes
(a), 30 minutes (b), 60 minutes (c).}
\label{Fig2}
\end{figure}

The microstructure of obtained samples was studied by SEM. We found that the
microstructure is highly sensitive to the heat-treatment conditions. The
analysis of SEM images showed us that heat-treatment increases grains sizes
due to their merging. This process is naturally accompanied by the growth of
contact areas between neighboring grains and thus to the better fitting
between them: grains become more compactly packed. If the heat-treatment
time is long enough, samples begin to recrystallize and intergrain
boundaries disappear.

Typical SEM images for CdS$_{0.2}$Se$_{0.8}$ films, which were heat-treated
at the temperature $823$ $K$, are presented in Fig. 2. Figs. 2(a), (b), and
(c) correspond to films which were heat-treated for 5, 30, and 60 minutes,
respectively. We found that the average grain size was around 3-12 $\mu m$
in all these cases with the tendency of this size to increase upon the
prolongation of the heat-treatment time, as clearly seen from Fig. 2. By its
structure, the film in Fig. 2(a) resembles a sand: relatively large spaces
exist between neighboring grains, these grains being connected to each other
only by rather small areas on their surfaces due to the rounded shape of
grains. At the same time, spacings between grains on Fig. 2(c) seem to
disappear, many of them now reduce to grooves. Fig. 2(b) corresponds to the
intermediate heat-treatment time, which leads to a compact packing of
grains; nevertheless narrow, long and deep gaps between them still exist.
Images of higher resolution, compared to the ones, presented here, support
these conclusions.

\begin{figure}[tbp]
\begin{center}
\includegraphics[width=0.5\textwidth]{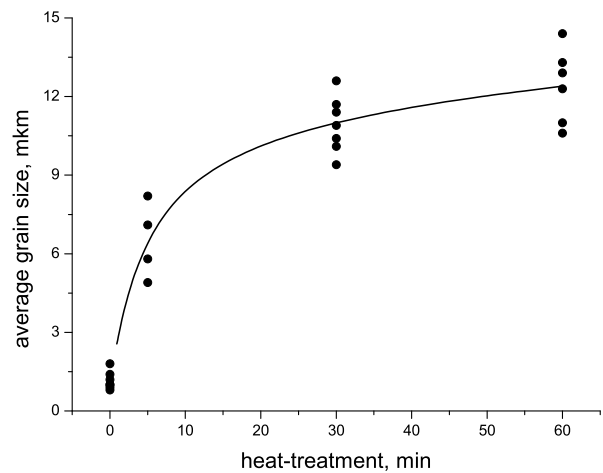}
\end{center}
\caption{The dependence of the average grain size on the annealing time for
24 samples heat-treated at the same temperature of 823 K.}
\label{Fig3}
\end{figure}

In order to demonstrate the tendency of the growth of the average grain size
with the increase of annealing time, we have taken 24 samples, which were
heat-treated at the same temperature of $823$ $K$, but during different
times. Then, we extracted the average grain size for each sample by
parforming the visual analysis of SEM images of these samples, each image
containing around 40-50 grains. The size of each grain was taken as an
average between its dimensions in the directions, where the grain has
smallest and largest dimensions. Our results are presented in Fig. 3, where
each dot corresponds to the particular sample, while the solid curve gives
the interpolation for the average grain size. It is clearly seen from Fig. 3
that the increase of annealing time indeed leads to the growth of grains.
This growth is however quite nonlinear with respect to the heat-treatment
time. Initially, heat-treatment leads to the rapid increase of the average
grain size and than this size tends to stabilize.

\begin{figure}[tbp]
\begin{center}
\centering
\subfigure{\includegraphics[width=0.9\textwidth]{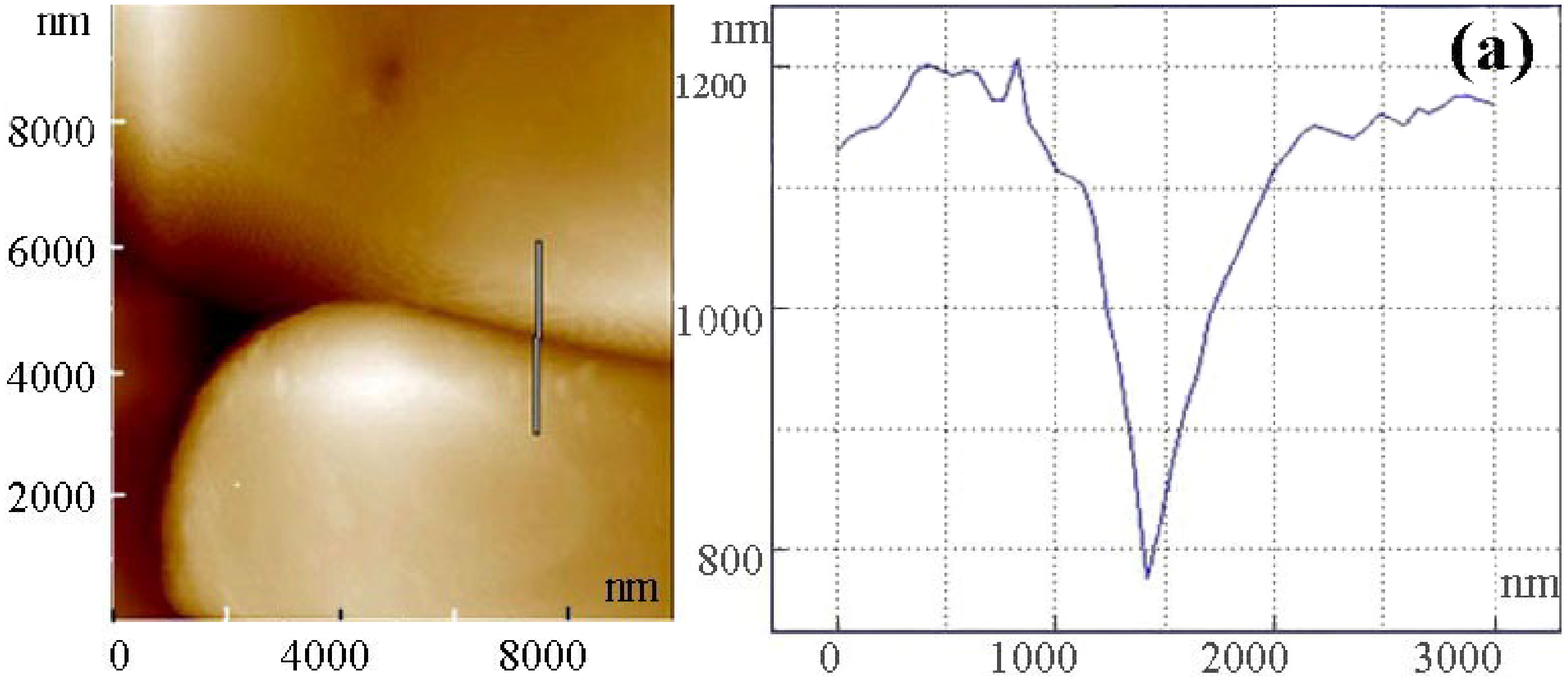}}\\[0pt]
\subfigure{\includegraphics[width=0.9\textwidth]{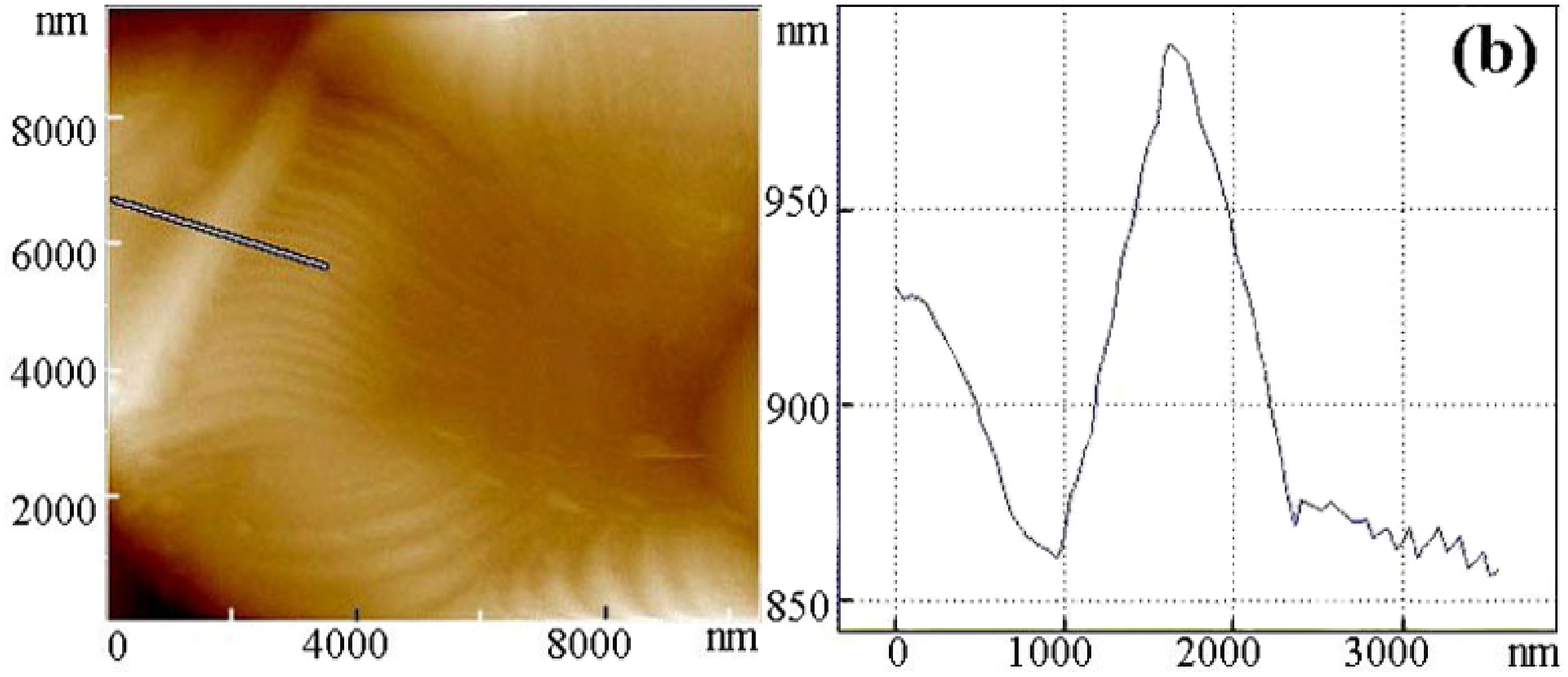}}
\end{center}
\caption{AFM images of intergrain boundaries for the sample, heat-treated
for 30 minutes (a) and for 60 minutes (b) at the same temperature of 823 K.
Left panels show the images themselves, right panels give the profiles of
landscapes along the lines marked by short lines in left panels.}
\label{Fig4}
\end{figure}

We also used a semicontact AFM with a thin cantilevers of whisker type in
order to study in a more detail the structure and geometry of boundaries
between grains. The typical radius of curvature for the whisker was 10 nm
and the typical height was 1 $\mu m$. Figs. 4 (a) and (b) show the AFM
images for the structure and profiles of typical grain boundaries in
samples, which were previously studied by SEM and which were fabricated
during the 30 and 60 minutes of heat-treatment, respectively, at the same
temperature of 823 K. The micro-relief of samples, obtained by 5 minutes of
heat-treatment, can not be investigated using this experimental approach,
since the surface landscape is very nonuniform in this case, which prevents
the use of a whisker.

It is seen from Fig. 3 that our expectations, obtained from the visual
analysis of SEM images, are basically correct: samples, which were sintered
during an intermediate time, contain narrow gaps, which do not disappear
completely during the fabrication process. These gaps become more and more
narrow when moving away from the film surface towards its interior regions.
Spacings between grains in films, which were heat-treated for a long time,
tend to disappear and they shrink into grooves on the film surface, see Fig.
4(b), which shows such a groove separating two regions of the grain with
different growth orientations.

\section{Experimental setup}

The experimental setup allowed us to study film's specific resistance by
spreading resistance method. For measurements of the resistance, we used
indium contacts, which were prepared by the transfer of melted indium from
Teflon plate under the pressure applied. The obtained contacts have a square
shape with one millimeter on side. The distance between contacts is also one
millimeter. Samples were placed in an isolated chamber that allows to study
the influence of various gaseous atmospheres (at pressures close to the
atmospheric pressure) and of vacuum. It was also possible to use streams of
gases for the same purposes. An incandescent lamp served as a source of
light. By changing the distance between the lamp and the film, we tuned the
intensity of light, which illuminated the samples. The film's resistance was
measured under the illumination up to 10$^{4}$ lux and at the temperature
varying from room temperature to 420 K.

The samples were placed on the stage, which was heated resistively.
Thermocouple was positioned inside the stage close to its surface, on which
the sample was situated. During the measurements, the temperature was
changing smoothly and nearly linear in time, with the rate of 1-1.5 K/min.
In order to control the accuracy of our measurements of temperature, we have
used an additional thermocouple, which was placed on the top of the sample.
The difference between the temperatures was within 0.1 K.

\section{Results and discussion}

\subsection{Photoresistance vs temperature}

As it was shown in Ref. \cite{Radio}, the lowest photoresistance at room
temperature was achieved for films, which were heat-treated during an
intermediate time, nearly 30 minutes, at heat-treatment temperature $823$ $K$%
. By photoresistance we mean sample's resistance under the illumination. In
this paper, we focus on temperature dependences of photoresistance $R(T)$
for samples characterized by different heat-treatment parameters. We expect
that these dependences can shed a light on the leading mechanism of charge
transfer. In total, for $R(T)$ measurements we used more than 30 samples.
The results were reproducible: for instance, for the photoresistance
maximum, the difference between different trials was within 5 \%.

Fig. 5 gives typical temperature dependences for the fixed illumination of
100 lux. It is clearly seen from this figure that the shape of all these
curves is rather peculiar: they consist of two segments demonstrating
opposite tendencies. At low temperatures, resistance grows with the increase
of temperature, while at high temperatures it decreases. The first type of
behavior is quite unusual for conventional semiconductors and insulators
(and common for metals), where charge transfer occurs by the simple thermal
activation of carriers through the forbidden gap. However, similar
temperature dependences of photoresistance were found in CdSe monocrystals 
\cite{Bube} with the "anomalous" part of $R(T)$ curve being localized in
nearly the same range of temperatures.

The non-monotonic behavior of $R(T)$ in CdSe monocrystals is explained by
the presence of centers in the forbidden gap belonging to two different
classes \cite{Bube,Rose}. These centers differ from each other by the fact
that centers of the first class have comparable capture cross sections both
for the free electrons and holes, while, for the centers of the second
class, capture cross section for holes is much larger. The latter centers
thus sensitize the photoconductor. By tuning light intensity or temperature,
one moves steady-state Fermi levels and changes the number of centers
contributing to the recombination. This gives rise to the superlinearity, as
well as to non-monotonic dependence of photoresistance on temperature. Since
quite peculiar temperature dependence of photoresistance found in our
granular films is similar to that in monocrystals \cite{Bube}, it is
reasonable to suggest that the leading mechanism of charge transfer under an
illumination is unique both in our samples and in monocrystals, while the
influence of internal boundaries is a less important factor. This factor,
however, is not negligible because it can provide an additional decrease of
photoresistance up to one order of magnitude in a certain range of
temperatures by varying annealing conditions, as seen from Fig. 5 (compare,
for instance, curves 1 and 3 near the temperature of 100 C).

We now discuss in a more detail how annealing conditions affects the
photoresistance. During the heat-treatment, the average grain size
increases, which results in the increase of contact areas between
neighboring grains and thus to the improvement of sample's connectivity. The
process of grains growth is very fast in the beginning (see Fig. 3) and
therefore the photoresistance of the sample heat-treated during 30 min.
(curve 2 in Fig. 5) is much lower than that of a sample, heat-treated during
5 min. (curve 1 in Fig. 5). However, heat-treatment also leads to the
intensive oxidation of grains surfaces, while the oxidation decreases
photoconductivity. Besides, merging of grains leads to the fact that
oxidized boundaries are now inside individual grains. Further increase of
heat-treatment time does not result in so intensive growth of grains sizes,
as seen from Fig. 3. The oxidation nevertheless still occurs. Moreover,
oxygen can diffuse from the bulk towards boundaries. Thus, oxidation starts
to lead to the increase of photoresistance with the prolongation of the
heat-treatment time (curves 2 and 3 in Fig. 5). In other words, an optimal
heat-treatment time depends on two competing mechanisms. Notice that the
existence of annealing temperatures, which maximize photosensitivity, was
reported in Ref. \cite{Garcia} for CdSe thin films obtained by a chemical
bath deposition technique.

Next, we discuss the position of photoresistance maximum along the
temperature axis in Fig. 5. It is seen from this figure that the temperature
corresponding to the photoresistance maximum is dependent on the
heat-treatment parameters, i.e., on microstructure. Increase of
heat-treatment time first leads to the increase of temperature at which this
maximum is achieved (curves 1 and 2). Such a behavior implies that the
contribution of the sensitizing centers is increased at this stage, so that
their effect is evident at curve 2 up to higher temperatures. This can be
due to improvement of crystallinity, as well as of intergrain boundaries,
where various defects, which can act as centers of the first class, should
be presented with higher concentration. However, further prolongation of
heat-treatment time results in the opposite shift of the corresponding
temperature (curves 2 and 3). This can be explained by oxidation, which
favors the appearance of additional centers of the first class.

It is worth noticing that the optimal heat-treatment time, which minimizes
the photoresistance, is strongly dependent on the temperature at which the
photoresistance is tested. While this optimal time was found to be
approximately 30 minutes for room temperature, it shifts towards 1 hour when
temperature is rising up to 160 C.

\begin{figure}[tbp]
\begin{center}
\includegraphics[width=0.5\textwidth]{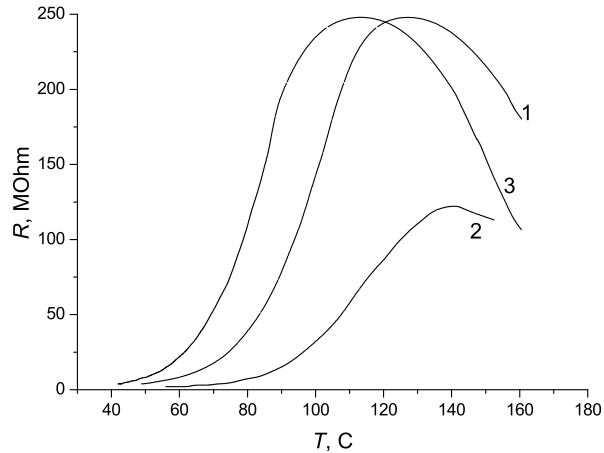}
\end{center}
\caption{Temperature dependences of film's photoresistance at fixed
illumination of 100 lux for samples with different times of heat-treatment
during their fabrication: 5 minutes (curve 1), 30 minutes (curve 2), 60
minutes (curve 3) and the same heat-treatment temperature of 823 K.}
\label{Fig5}
\end{figure}

An important electrophysical characteristic of photoconductive materials is
a dark resistance as a function of temperature. In Fig. 6, we plot the
dependence of resistance without illumination on temperature for the same
three samples, addressed in Fig. 5. We see that the logarithm of resistance
depends linearly on inverse temperature, in accordance with the Arrhenius
law. From the slopes of these dependences we can extract activation
energies. We have found that these energies are nearly the same for all the
three samples and are equal to $2.0$ eV with the error $\pm $0.1 eV. The
dark resistance of optimally-prepared films is the highest one, while the
resistance of the sample, heat-treated during 60 min., is the lowest one.
This can be understood by noting that the latter case corresponds to the
very well connectivity of the sample due to merging of individual grains. At
the same time, heat-treatment also leads to oxidation of grain boundaries,
which suppresses charge transfer over intergrain barriers. The competition
between the two mechanisms, in the case of absence of illumination, leads to
the domination of the oxidation effect for the case of optimally-prepared
films in contrast to the resistance under the illumination.

\begin{figure}[tbp]
\begin{center}
\includegraphics[width=0.5\textwidth]{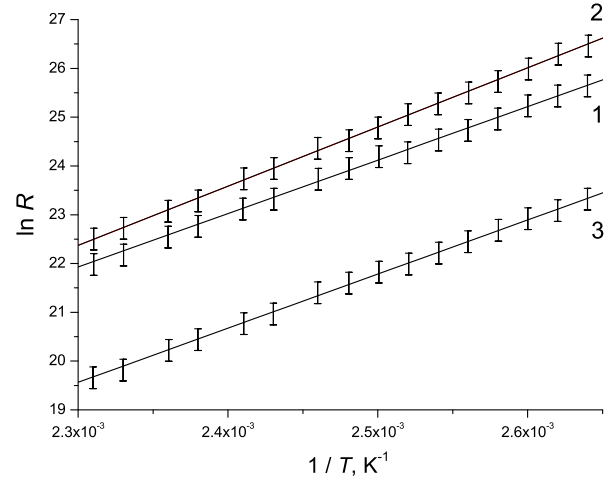}
\end{center}
\caption{Temperature dependences of film's resistance without illumination
for samples with different times of heat-treatment during their fabrication:
5 minutes (curve 1), 30 minutes (curve 2), 60 minutes (curve 3), and the
same heat-treatment temperature of 823 K.}
\label{Fig6}
\end{figure}

The value of the activation energy is in agreement with the width of the
forbidden gap of CdS$_{0.2}$Se$_{0.8}$ and the measured photocurrent
spectra, one of which is presented in Fig. 7 (for the sample, heat-treated
during 5 minutes at $T=823$ K). Measurements were performed using a scanning
spectrometer at room temperature. Spectral range was from 400 to 900 nm. The
spectra were decomposed into separate lines (Gauss or Lorentz functions)
with the confidence probability of 95 \%. The smallest discrepancy value was
used as a criterion of decomposition. All the measured spectra contain three
components, which correspond to energies 2.0 eV (curve 1), 1.7 eV (curve 2),
and 1.6 eV (curve 3). The first one is due to direct transitions in the
forbidden gap, while the two others are linked to the transitions with
participation of centers in this gap. Curve 4 in Fig. 7 gives the sum of the
three contributions, and dots correspond to experimental data.

\begin{figure}[tbp]
\begin{center}
\includegraphics[width=0.5\textwidth]{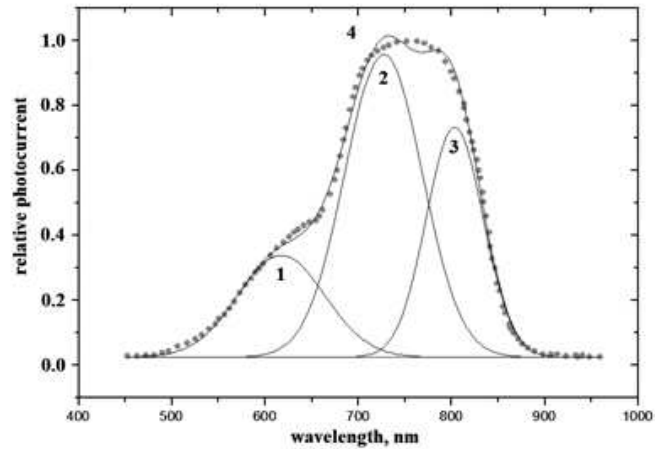}
\end{center}
\caption{The dependence of the relative photocurrent intensity on the
wavelength of illumination for the sample, heat-treated during 5 minutes at $%
T=823$ K.}
\label{Fig7}
\end{figure}

Note that we have also studied the response of the sample to the rectangular
illumination impulse with the amplitude of nearly 200 lux with the width of
the order of 10$^{-5}$ s. We have found that the typical photoresponse time
was of the order of 10$^{-6}$ s.

\subsection{Influence of ambient atmosphere}

In order to better understand the role of oxygen and water, which are
absorbed/chemiapsorbed at the surfaces, we performed experimental
investigations of photoresistance in helium atmosphere, in vacuum, and also
in the stream of helium, after keeping samples in these environment for 1
hour. These conditions diminish oxygen and water absorption at intergrain
boundaries, as well as at the film's surface. Of course, an oxide layer
cannot be removed by this method, since this layer essentially appears
during the heat-treatment with a quasi-free air access.

Fig. 5 presents the typical results for $R(T)$\ dependence at constant
illumination of 1500 lux in the atmosphere of air, helium, and in the vacuum
for the samples heat-treated for 30 minutes at temperature 823 K. Firstly,
shape of all these curves is not dependent on the atmosphere, which
evidences that non-monotonic $R(T)$\ dependence cannot be due to
oxidation/deoxidation processes during sample's heating/cooling: This is
consistent with the supposition that the dominant mechanism of charge
transfer is the same as in monocrystals. Secondly, it is seen from Fig. 5
that the lowest photoresistance is achieved in the vacuum, i.e., when oxygen
and water absorption is suppressed, while the highest photoresistance shows
up in the air atmosphere.

The influence of absorbtion is not very strong at low temperatures, i.e., at
the "anomalous" parts of $R(T)$ curves, since there absolute values of
photoresistance for different curves vary only within 30 \%. However, at
higher temperatures, when approaching the thermoactivation part of $R(T)$
curves, an atmosphere starts to play an important role. For instance, the
resistance of the sample in vacuum (curve 3) is approximately two times
lower than that in the air (curve 1) at $T=130$\ $C$. The resistance of the
sample in helium atmosphere (curve 2) is lower than that in the air (curve
1), but also higher than the resistance in vacuum (curve 3). This can be
attributed to the weaker desorption of air and water in helium environment
compared to the vacuum.

\begin{figure}[tbp]
\begin{center}
\includegraphics[width=0.5\textwidth]{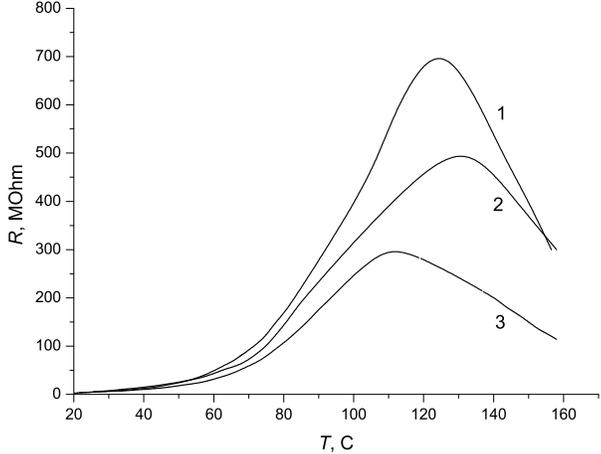}
\end{center}
\caption{Temperature dependences of film's photoresistance at fixed
illumination of 1500 lux in different environments: in air (curve 1), in
helium (curve 2), in vacuum (curve 3). The sample was heat-treated during 30
minutes at the temperature of 823 K.}
\label{Fig8}
\end{figure}

We finally notice that different studies of CdS and CdSe films performed in
the past provided different results for the temperature dependence of
photoresistance. For instance, in Ref. \cite{Gupta}, where photoresistance
of CdS$_{1-x}$Se$_{x}$ polycrystalline films was investigated, an
"anomalous" part of $R(T)$ curve was not found, the whole dependence being
of "thermoactivation" nature. Ref. \cite{Pillai1} deals with CdSSe(Cu) in
silicone resine binder layers, which also demonstrates simple monotonic
behavior. At the same time, non-monotonic $R(T)$ dependences, similar to the
ones presented here, were reported in Ref. \cite{Pillai2} for CdS$_{1-x}$Se$%
_{x}$ sintered layers and in Ref. \cite{Aneva} for CdSe thin films prepared
by thermal vacuum evaporation.

\section{Conclusions}

Highly photosensitive granular CdS$_{0.2}$Se$_{0.8}$ films were fabricated
by the screen printing method. X-ray analysis has shown that grains consist
of a solid CdS-CdSe solution. We measured temperature dependences of
photoresistance and found that they have a peculiar non-monotonic shape,
which is practically identical with that for CdSe monocrystals, known from
literature. Namely, photoresistance first increases with the increase of
temperature and then it starts to decrease. We therefore conclude that, in
these granular films, the leading mechanism of charge transfer under the
illumination is the same as in monocrystals, i.e., based on the presence of
two kinds of centers in the forbidden gap.

Influence of intergrain boundaries is however not negligible. This follows
from the fact that photoresistance also depends on the film's
microstructure, which can be changed by tuning the time and temperature of
heat-treatment during the process of sample's fabrication. The longer
heat-treatment time, the larger grains and the more compactly they are
packed. The lowest photoresistance at room temperatures was found for films,
heat-treated for an intermediate time, in which gaps between grains still
exist. This "optimal" heat-treatment time, however, strongly depends on the
temperature, at which photoresistance is probed, namely, it decreases with
the increase of temperature. For a certain range of temperatures, one can
suppress photoresistance as much as up to one order in magnitude by only
varying film's internal microstructure.

An explanation for the optimal heat-treatment time was suggested in terms of
the competition between the two mechanisms: longer heat-treatment improves
film's connectivity, but also results in additional oxidation. The optimal
heat-treatment time thus depends on the interplay between these two factors.

To better understand the effect of grain boundaries, we performed additional
measurements of photoresistance in vacuum and also in helium atmosphere,
which have shown that oxygen and water absorbtion/chemisorption at grain
boundaries increases photoresistance.

\section{Acknowledgements}

The authors are grateful to L. N. Survilo for the help in the preparation of
samples and to D. Strateychuk for making films images. W. V. P. acknowledges
the support from the "Dynasty Foundation" for young scientists.

\end{document}